\documentclass[traditabstract]{aa} 
\usepackage{graphicx,natbib,psfig,amssymb,multirow,sidecap} 
\newcommand{\ltsima} {$\; \buildrel < \over \sim \;$} 
\newcommand{\gtsima} {$\; \buildrel > \over \sim \;$} 
\newcommand{\lta} {\lower.5ex\hbox{\ltsima}} 
\newcommand{\gta} {\lower.5ex\hbox{\gtsima}} 
\newcommand{\Ha} {H$\alpha$}
\newcommand{\Hb} {H$\beta$}
\newcommand{\ergscm}{\>{\rm erg}\,{\rm s}^{-1}\,{\rm cm}^{-2}}
\newcommand{\kms}{$\rm{\,km \,s}^{-1}$}
\newcommand{\forb}[2]{\mbox{$[{\rm #1\, #2}]$}}
\newcommand{\oiii}{\forb{O}{III}}
\begin{document} 
\title{Emission lines in early-type galaxies: \\
active nuclei or stars?}
  
\titlerunning{Emission lines in early-type galaxies}

\authorrunning{A. Capetti \& R.D. Baldi}
  
\author{Alessandro Capetti \inst{1} \and
Ranieri D. Baldi
\inst{1,2}} 
\offprints{A. Capetti}  
\institute{INAF - Osservatorio Astronomico di Torino, Via
  Osservatorio 20, I-10025 Pino Torinese, Italy \and
Space Telescope Science Institute, 3700 San
  Martin Drive, Baltimore, MD 21218, U.S.A.\\
\email{capetti@oato.inaf.it}
\email{baldi@stsci.edu}}

\date{}  
   
\abstract{We selected 27244 nearby, red, giant early-type galaxies (RGEs) from
  the Sloan Digital Sky Survey (SDSS). In a large fraction (53 \%) of their
  spectra the [O~III]$\lambda5007$ emission line is detected, with an
  equivalent width (EW) distribution strongly clustered around $\sim$0.75
  \AA. The vast majority of those RGEs for which it is possible to derive
  emission line ratios (amounting to about half of the sample) show values
  characteristic of LINERs.

  The close connection between emission lines and stellar continuum 
    points to stellar processes as the most likely source of the bulk of the
    ionizing photons in RGEs, rather than active nuclei. In particular, the
    observed EW and optical line ratios are consistent with the predictions of
    models in which the photoionization comes from to hot evolved stars. Shocks
    driven by supernovae or stellar ejecta might also contribute to the
    ionization budget.

    A minority, $\sim$4\%, of the galaxies show emission lines with an
    equivalent that is width a factor of $\gtrsim 2$ greater than the sample
    median. Only among them are Seyfert-like spectra found. Furthermore, 40\%
    of this subgroup have a radio counterpart, compared to $\sim$6\% of the
    rest of the sample. These characteristics argue in favor of an AGN origin
    for their emission lines.

    Emission lines diagnostic diagrams do not reveal a distinction between the
    AGN subset and the other members of the sample, and consequently they are
    not a useful tool for establishing the dominant source of the ionizing
    photons, which is better predicted by the EW of the emission lines.

  }

\maketitle

\section{Introduction}

The presence of ionized gas in early-type galaxies, part of a complex
multiphase interstellar medium (ISM), has now been firmly established. Among
other issues, the source of ionization that powers the observed emission lines
is still a matter of debate. Several alternatives have been advanced,
ascribing it to active galactic nuclei (AGN), to hot evolved stars, and to
shocks (e.g. \citealt{heckman80,binette94,dopita95}). A better understanding
of this phenomenon is important for several reasons, e.g. in the contexts of
the origin and evolution of early-type galaxies and of their ISM and 
of a correct census and demography of active galaxies.

In this paper we explore this issue by studying homogeneous sample of
galaxies, restricted to a rather narrow range of galaxy properties. We isolated
nearby, quiescent (from the point of view of star formation), massive
early-type galaxies from the SDSS \citep{york00}. More specifically, from the
main sample of $\sim$800000 galaxies with spectra available from the SDSS,
Data Release (DR) 7, we used the MPA-JHU DR7 release of spectrum measurements,
available at {\sl http://www.mpa-garching.mpg.de/SDSS/DR7/}, to select all
sources with the following spectroscopic criteria: 1) redshift between 0.01
and 0.1, 2) Ca break strength $D_n(4000)>1.7$, and 3) stellar velocity
dispersion $\sigma_* > 156$ \kms. If adopting the scaling relation between mass
and velocity dispersion this corresponds to galaxies with $M_* \gtrsim 5
\times 10^{10} M_{\sun}$, \citep{hyde09}. We only retain early-type
galaxies, i.e. objects with a concentration index (the ratio of the radii
including 90\% and 50\% of the $r$ band light) $C_{r}\geq 2.86$
\citep{nakamura03,shen03}. This selection yields 27244 red, giant, early-type
galaxies (hereafter RGEs).

\begin{figure*}
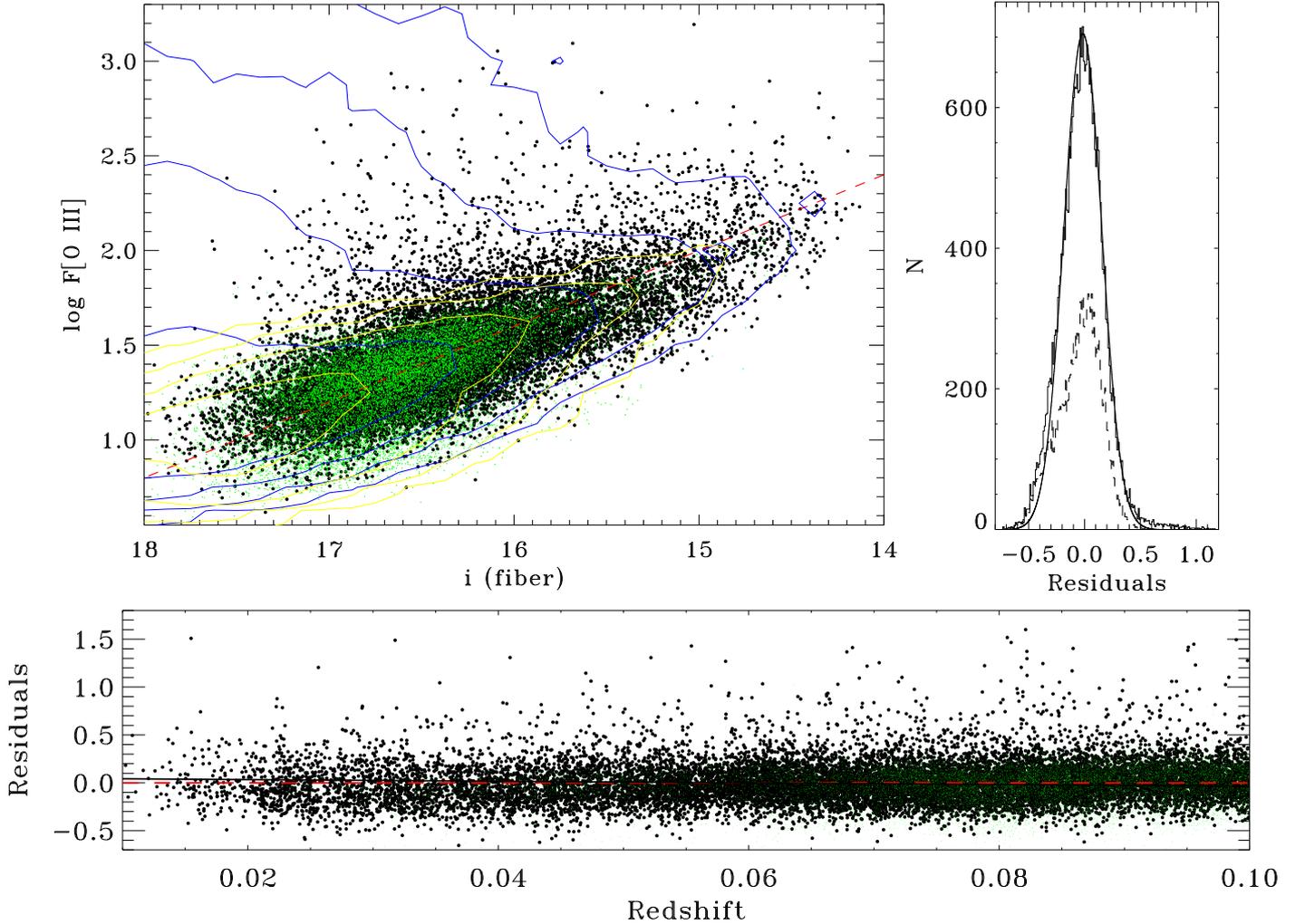

\centerline{
\includegraphics[scale=0.75,angle=0]{16388f1a.epsi}
\medskip
\includegraphics[scale=0.76,angle=0]{16388f1b.epsi}
}
\includegraphics[scale=1.15,angle=0]{16388f1c.epsi}
\caption{Left panel: logarithm of the [O~III] emission line flux (in units of
  $10^{-17}\ergscm$) versus the k-corrected $i$ band magnitude within the
  SDSS fiber, both quantities corrected for galactic absorption.  Green dots
  mark upper limits in line flux. The dashed red line corresponds to a
  constant ratio between the two quantities. Contours (blue for the objects
  with an [O~III] detection, yellow for upper limits) represent the
  iso-densities of DR7 galaxies in the same redshift range; levels are in
  geometric sequence with a common ratio of 4. Right panel: histogram of the
  residuals from the median line. The dashed histogram is the contribution of
  upper limits. The dispersion of the distribution is 0.18 dex. Bottom:
  residuals from the median vs redshift.}
\label{riga}
\end{figure*}

\section{Emission lines and stellar continuum}
\label{el}

We considered the [O~III]$\lambda5007$ emission line flux measured after
subtracting a starlight template (see \citealt{kauffmann03} for a detailed
description of the method used for the continuum subtraction). The [O~III]
line is detected at a significance higher than 3$\sigma$ in $\sim$ 53\% of
RGEs of the sample. Figure\ref{riga} compares the [O~III] flux with the
k-corrected $i$ band magnitude within the SDSS fiber, with both quantities
corrected for galactic absorption. A strong connection emerges, and the vast
majority of the objects are clustered in a narrow stripe. We used the
algorithms proposed by \citet{dempster77} and \citet{buckley79}, implemented
as {\sl emmethod} and {\sl buckleyjames} in IRAF, to deal with censored data
in order to derive the best-fitting logarithmic slope of the $F_{\rm[O~III]}$
versus $i$ relation. The resulting values are $0.41\pm0.01$ and $0.40\pm0.01$,
respectively, both consistent with a constant ratio between emission line and
stellar continuum.

In the right panel we show the histogram of the residuals from the median line
that is reproduced well by a Gaussian distribution with a dispersion of only
0.18 dex (similar results, although with slightly broader distributions, are
obtained using different lines and/or continuum bands). Nonetheless, there are
outliers with a line excess, R\oiii\footnote{Defined as the ratio between the
  \oiii\ flux of a galaxy with respect to its median value at a given
  magnitude.}, of up to 100.  Setting a threshold at R\oiii = 5, we find 0.5\%
outliers, a fraction that increases to 1.2\% (3.5\%) when lowering the limit
to 3 (2).

The EW of the [O~III] line, estimated by considering the continuum level 200
pixels around the line, has a median value of $\sim 0.75$ \AA\ including
upper limits in the analysis) and the EW dispersion is of 0.20 dex; 95
\% of RGEs with an \oiii\ detection have 0.3 $<$ EW $<$ 1.7 \AA.

Contours in Figure\ref{riga} represent the iso-densities of the $\sim$340000
DR7 galaxies in the same redshift range, $0.01<z<0.1$, but without any further
selection of the spectro-photometric properties. The general galaxies
population shows a concentration in the same locus as covered by the RGEs, but
also a substantial population extending toward high line excesses ($\sim$ 30\%
of the sample has an excess of R\oiii $>$3).

We looked for trends between the residuals and the spectrophotometric
parameters used for the sample selection (namely redshift, $D_n(4000)$,
concentration index, and velocity dispersion), but we failed to find any
statistically significant link. In particular, there are no differences
between RGEs at low and high redshifts: the best linear fit of the residuals
against redshift indicates that objects at z=0.01 have a ratio between line
and continuum only $(20\pm11)$\% higher than those at z=0.1.

\begin{figure*}
\centerline{
\includegraphics[scale=1.,angle=0]{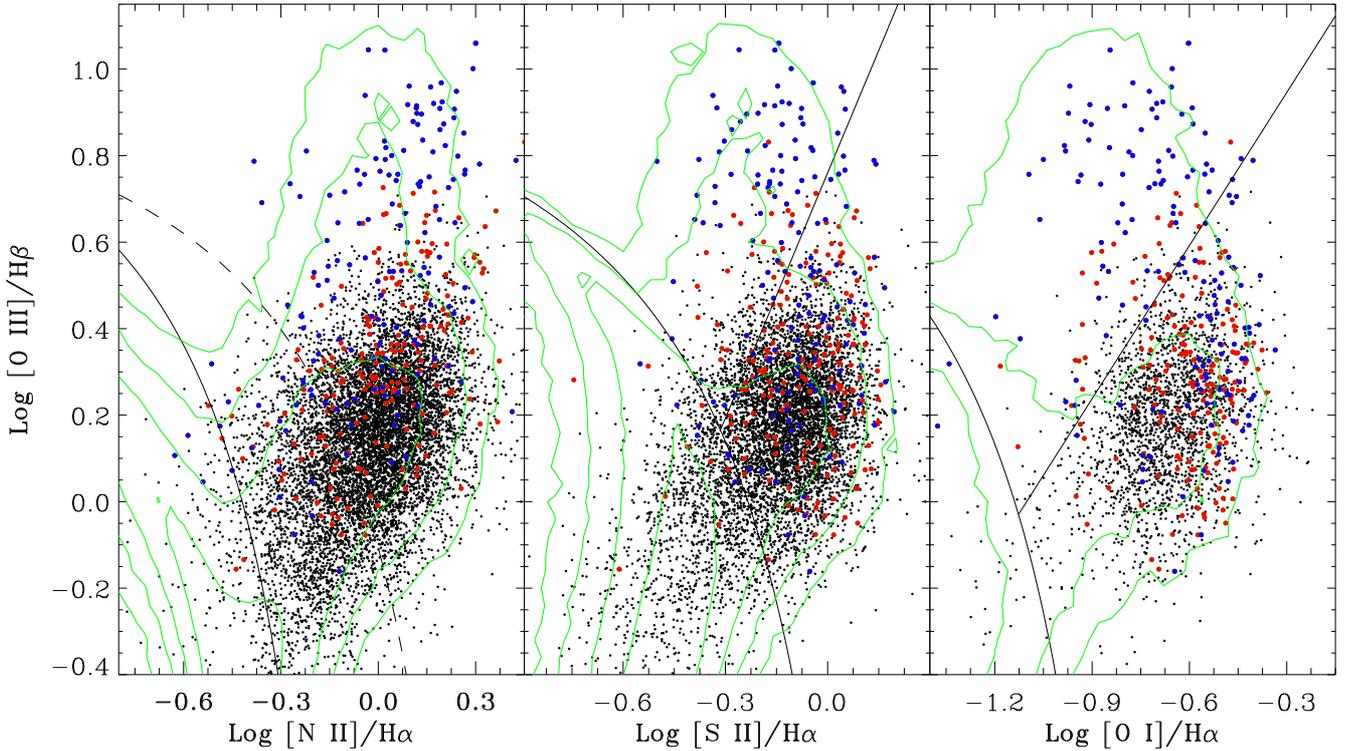}
}
\caption{Spectroscopic diagnostic diagrams for RGEs.  The solid lines are from
  \citet{kewley06} and separate star-forming galaxies, LINER, and Seyfert; in
  the first panel the region between the two curves is populated by the
  composite galaxies. Blue dots mark the ``strong'' outliers (objects with a
  line excess with respect to the median value R\oiii$>$5) in Figure\ref{riga},
  and red dots the ``weak'' outliers (3$<$R\oiii$<$5). Contours represent
  the iso-densities of all DR7 emission line galaxies.}
\label{diag}
\end{figure*}

\begin{figure}
\centerline{
\includegraphics[scale=0.5,angle=0]{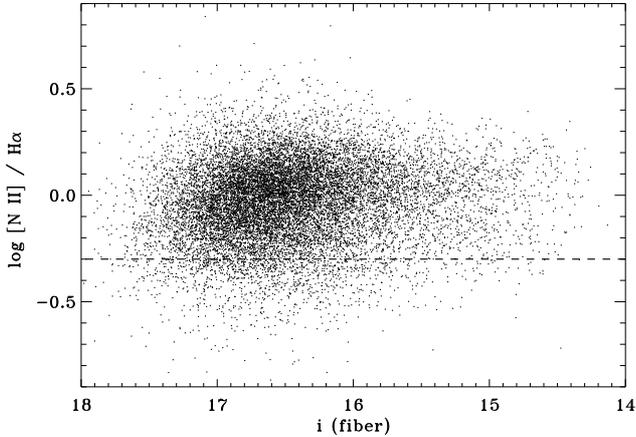}
}
\caption{[N~II]/\Ha\ ratio for all sources with both lines detected. Emission
  lines of the objects above the horizontal dashed line are unlikely to be
  powered by star formation.}
\label{n2ha}
\end{figure}

\section{Spectroscopic diagnostic diagrams}

Fig.~\ref{diag} shows the location in the spectroscopic diagnostic diagrams
(e.g. \citealt{heckman80,baldwin81,veilleux87,kewley06}) for the RGEs that
have all the relevant emission lines detected at SNR$>$3 separately for each
diagram. Starting from the left side, the percentages with respect to the
whole sample are of 29, 23, and 8\% in the three diagrams, respectively. The
vast majority of the objects fall in the LINERs region, while the Seyfert and
star-forming regions are scarcely populated.

Considering the properties of outliers first, `strong' outliers (i.e. the
objects with R\oiii\ $>$ 5) represent essentially all of the Seyferts;
nonetheless, most of them are LINERs. `Weak' outliers (with $3 <$ R\oiii $< 5$)
do not differ significantly, on the one hand, from the `stronger' ones (but
there are fewer Seyfert among them), but on the other, they cannot be
readily separated from the bulk of the RGEs population.

We now turn our attention to the objects that have no optical classification
because at least one of the key emission lines is undetected. This subsample
is clearly very important, because it is composed of $\sim$ 3/4 of the
galaxies. Nonetheless, [N~II] and \Ha\ are both detected in $\sim$50\% of the
galaxies and this allows us to derive at least a tentative classification. In
fact, log([N~II]/\Ha) $> -0.3$ in 90 \% of them (see Figure\ref{n2ha}), a
threshold above which no star-forming galaxies are found. On the other hand,
Seyfert galaxies usually have bright emission lines, thence they are expected
to be all properly cataloged by the diagnostic diagrams in
Fig.~\ref{diag}. This leads to the conclusion that these galaxies, amounting
to about half of the RGEs, are generally LINERs.

\section{Error budget of the emission line fluxes}

\begin{figure}
\centerline{
\includegraphics[scale=0.6,angle=0]{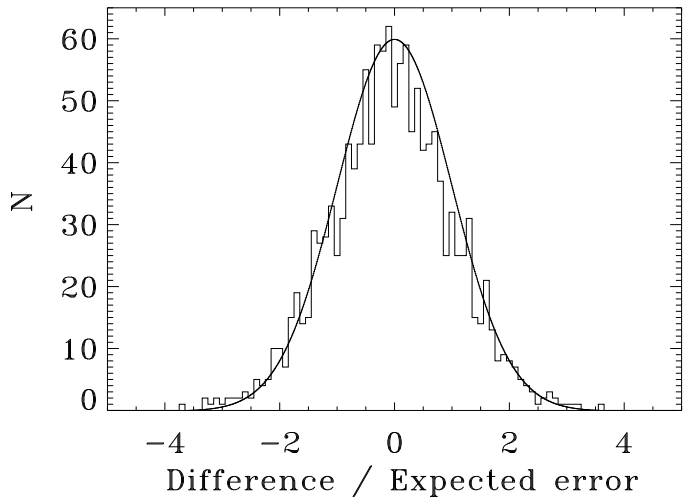}
}
\caption{Difference between pairs of duplicated [O~III] flux measurements
  divided by the error estimated propagating the individual uncertainties.
  The solid line is a Gaussian distribution with a dispersion of 1.}
\label{dupl}
\end{figure}

Since we are studying emission lines generally of very low EW, it is important
to assess the reliability of such measurements and of the error
estimates. Two sources of uncertainties should be considered, the first
  related to statistical errors, the second to the effects of starlight
  modeling.

  We deal with statistical uncertainties by studying the galaxies with
  multiple SDSS spectroscopic observations. There are 3103 such objects: 45\%
  of them have the [O~III] line detected in both datasets, 35\% are upper
  limits in both observations, while 20\% change from detected to undetected
  (or vice-versa). Considering the objects detected in both observations, the
  differences between the independent measurements closely follow a Gaussian
  distribution with a dispersion equal to what is expected based on the
  uncertainties in each flux estimate (see Fig ~\ref{dupl}). We conclude that
  the error assessment is robust and that most (85\%) of the [O~III]
  detections are confirmed by multiple observations.

  In addition to the statistical uncertainties, we must also consider the
  effects related to the accuracy of the subtraction of the stellar
  emission. As discussed in detail by \citet{annibali10}, because of the
  degeneracy between age, metallicity, and extinction, fits to the continuum
  emission of similarly good quality can produce significantly different
  residual spectra. This is particular relevant for the \Hb\ line (and to a
  lesser extent for \Ha) which is superimposed on the stellar Balmer
  absorption feature. The method proposed by \citet{annibali10} to cope with
  this problem in their study of a sample of nearby ETG is to explore the
  changes in the emission lines fluxes related to the various stellar
  templates, limiting it to those that yield a fit with $\chi^2<2$. The
  adopted emission line flux is a weighted average of those derived from all
  acceptable fits, while its error is a weighted value of the dispersion
  between the individual estimates.

  The very large number of SDSS objects considered prevents a detailed
  analysis of the error budget considered by these authors. However, this
  provides us with guidance on the typical uncertainties that should be
  associated with the line measurements due to a template mismatch. From their
  published values for their smaller synthetic aperture (comparable in size
  with the SDSS fiber), we estimated that the median errors in the EW for \Hb,
  [O III], and \Ha\ are 0.16, 0.08, and 0.08 \AA, respectively. We adopt these
  reference values also for the SDSS sources. The errors in EW have been been
  converted into a line flux error and then added quadratically to the
  uncertainties provided by the SDSS database.

  Our results for the connection between \oiii\ flux and $z$ magnitude are only
  marginally changed by adopting this more conservative error treatment. The
  fraction of detected sources decreases from 53 to 50\%, but the slope of the
  relation remains unchanged. The number of outliers is not affected.

  We then reconsidered the diagnostic diagrams, excluding all sources that do
  not meet the 3$\sigma$ criterion with the revised errors. The fraction of
  sources in the various diagrams decreases by 25 to 40 \%, but their visual
  appearance is essentially preserved. In order to quantify more subtle
  differences, we estimated the changes of median values and dispersions of
  the various ratios. Not surprisingly log$\,$\oiii/\Hb\ is the more affected
  ratio, because it changes from 0.15 (with an rms of 0.20) to 0.09, (rms =
  0.18 dex). This is most likely due to the preferred exclusion of galaxies
  with low \Hb\ EW. Nonetheless, the result that the vast majority of the
  sources are LINERs is confirmed.

  Finally, we explored how crucial the precise value of the adopted errors
  on EW is. Doubling the reference values, the number of sources in the
  diagnostic diagrams keeps decreasing, but their location is effectively
  unchanged, with a median of log$\,$\oiii/\Hb\ = 0.05.

\section{Discussion}

Most nearby, red, giant early-type galaxies are emission line galaxies. In
particular, in 53\% of their SDSS spectra, the [O~III]$\lambda5007$ line is
detected. The majority of the objects for which this analysis was
possible show emission line ratios characteristic of LINERs.

The [O~III] flux shows a strong correlation with the flux measured within the
3$\arcsec$ SDSS fiber in the $i$ band. The $i$ band magnitude (being less
affected by uncertainties in the k-correction, absorption, and by differences
in the stellar population) is a good estimator of the stellar mass within the
fiber. Thus the ratio between lines and stellar mass is essentially constant
(showing a dispersion of only 0.18 dex) while both quantities vary by a factor
of $\sim$ 30. Furthermore, there is only a marginal trend for a change of EW
with redshift: objects at z=0.01 have a ratio between line and continuum
$(20\pm11)$\% higher than those at z=0.1 despite the change in the size of the
region covered by the SDSS fiber from $\sim$0.6 to 5.5 kpc. 

These two results are not straightforward for explaining if the emission lines
are powered by an AGN. In fact, they require a fine tuning between the
strength of the nuclear ionizing field, the spatial distribution of the
emission lines, and the stellar mass. A general link between AGN activity and
the {\sl total} stellar mass can be envisaged, via the constant ratio between
the mass of stars and of the black hole \citep{marconi03}, while we find a
connection with the stellar mass covered by the SDSS fiber. Furthermore, also
the accretion rate, the second parameter that together with the black hole
mass sets the radiative output of the AGN, should be closely linked with the
amount of stars within the central 3$\arcsec$ of each given galaxy.

Conversely, these findings strongly suggest that processes related to the
  stellar population are at the origin of the observed emission lines.

  As argued by several authors
  (e.g. \citealt{binette94,macchetto96,stasinska08,sarzi10}), hot evolved
  stars, such as post-asymptotic giant branch stars (pAGB) and white dwarves
  (WD), can produce a substantial diffuse field of ionizing photons. More
  quantitatively, the observed median EW for \Ha\ in the sample considered
  here is EW$_{\rm H\alpha} \sim 0.8$ \AA, well within the range $0.6-1.7$
  \AA\ predicted by \citet{binette94} for a stellar population of age
  $\sim10^{10}$ years, assuming that the cold gas intercepts all the ionizing
  photons. \citet{stasinska08} have explored photoionization models in which
  the Lyman continuum radiation is directly estimated from a stellar
  population analysis, and find that, by varying the metallicity and
  ionization parameter $U$, the resulting emission line ratios cover the whole
  region typical of LINERs in the spectroscopic diagnostic diagrams. In
  particular, the location of RGEs is reproduced well by models with
  $0.5\lesssim Z_*/Z_{\sun}\lesssim 2$ and log $U \sim -3.7$.

Shocks can represent an additional source of ionization. It has been
  shown by, e.g., \citet{allen08} that they can produce emission lines with
  ratios mimicking those of active nuclei and of LINERs, in particular. The
  connection between stellar mass and emission line flux suggests that such
  shocks must be driven by the stellar population, as in the case of
  supernovae or of stellar ejecta, rather than by bars of triaxial
  perturbations.

We conclude that the properties of emission lines in RGEs can in general be
satisfactorily accounted for by photoionization related to hot evolved
  stars or by shocks driven by the stellar processes (but below we discuss the
  few and important exceptions of likely AGN dominated objects). The narrow
spread of observed EW, very difficult to explain in an AGN framework, set
strong constraints even when  adopting a ``stellar'' interpretation.  For example,
in the scenario of lines powered by an evolved stellar population, the
apparently stringent requirement that the cold gas must intercept essentially
all of the ionizing photons, becomes probably instrumental in producing a
narrow distribution of lines EW. In fact this effectively removes the
fundamental degree of freedom related to the ISM optical depth to ionizing
photons. The selection of galaxies within a rather narrow range of the various
spectro-photometric parameters space is also a key ingredient in reducing the
scatter between F\oiii\ and $i$ magnitude (see Figure\ref{riga}).

These results were obtained by focusing on RGEs, but the consequences are
probably more far reaching. In fact, in the F\oiii\ vs. $i$ plane a strong
concentration is also seen for the general population of galaxies, in the same
region covered by RGEs. This suggests a wider application, possibly extending
to less massive early-type galaxies and to the bulges of spiral galaxies.

Nonetheless, there are outliers from the general link between stellar and line
emission. Only among these sources we found Seyfert-like spectra. Furthermore,
a strong connection emerges between the excess in line emission and the
presence of a nuclear radio source, a characteristic signature of an AGN. Our
selection criteria isolate galaxies whose properties are closely matched to
the typical hosts of low-luminosity radio-loud AGN \citep{baldi10b}.
Considering the outliers with R\oiii $>5$ (3), 49\% (45\%) have a radio
counterpart in the FIRST\footnote{Faint Images of the Radio Sky at Twenty
  centimeters survey \citep{becker95}} catalog within 2$\arcsec$ from the
optical position. In the rest of the sample, only 6\% of the sources have a
radio counterpart. The connection between outliers and radiosources is
preserved even at very small line excesses, since 31\% of those with $2 <
$R\oiii $<3$ and 17\% of the objects with an excess by only a factor of 1.5 to
2, are associated with a FIRST source. Thus, the galaxies that exceed the
rather narrow range of line emission EW set by the stellar processes are very
often associated with an active nucleus. The observed line excess is likely to
stem from the additional source of ionizing photons represented by the
AGN. The rather loose relation between radio power and line luminosity seen in
these low-luminosity AGN (see the results presented by \citealt{baldi10b})
might explain the incomplete coincidence between line excess and the presence
of a radio source and, on the other hand, the reason not all radio sources
produce a detectable increase in the line EW.

Apparently, the location in the spectroscopic diagnostic diagrams is not
sufficient to establish the dominant source of ionizing photons, which,
conversely, is best predicted by the EW of the emission lines. In fact, the
likely AGN show line ratios that are essentially indistinguishable (leaving
aside the few Seyfert-like objects) from those probably dominated by stellar
processes.  As already pointed out by \citet{stasinska08}, this has important
ramifications where the census and the properties of AGN are concerned. To
isolate and explore the properties of genuine low-luminosity active galaxies,
spectroscopic observations obtained with a smaller aperture are needed to
increase the contrast between AGN and stellar induced line emission.

\acknowledgements

We thank the David J. Axon and anonymous referee for comments that improved
the clarity of this paper.

\end{document}